\documentclass[%
aip,
apl,
amsmath,amssymb,
preprint,%
floatfix,
]{revtex4-2}

\usepackage[dvipdfmx]{graphicx}%
\usepackage{dcolumn}%
\usepackage{bm}%

\usepackage{siunitx}
\usepackage{xcolor}
\usepackage{comment}
\usepackage{enumerate}
\usepackage[version=4]{mhchem}
\usepackage{color}

\makeatletter
\def\@email#1#2{%
 \endgroup
 \patchcmd{\titleblock@produce}
 {\frontmatter@RRAPformat}
 {\frontmatter@RRAPformat{\produce@RRAP{*#1\href{mailto:#2}{#2}}}\frontmatter@RRAPformat}
 {}{}
}%
\makeatother
\begin{document}

\preprint{}

\title{One-step Fabrication of Sharp Platinum/Iridium Tips via Amplitude-Modulated Alternating-Current Electropolishing}

\author{Yuto Nishiwaki}
\author{Toru Utsunomiya}
\author{Shu Kurokawa}
\author{Takashi Ichii*}%
 \email{ichii.takashi.2m@kyoto-u.ac.jp}
\affiliation{ 
 Department of Materials Science and Engineering, Kyoto University, Yoshida Honmachi, Sakyo, Kyoto, 606-8501, Japan. %
}%

\date{\today}

\begin{abstract} 
  
The platinum/iridium (Pt/Ir) alloy tip for scanning probe microscopy (SPM) was fabricated by amplitude-modulated alternating-current (AC) electropolishing. The clean tips with a radius of curvature less than \SI{100}{nm} were reproducibly obtained by applying the sinusoidal voltage in the frequency ($f_0$) of $\SI{900}{Hz}\leq f_0\leq\SI{1500}{Hz}$ with amplitude modulation by the sinusoidal wave in the modulation frequency ($f_s$) of $f_s=0.1f_0$ in \ce{CaCl2}/\ce{H2O}/acetone solution. The analyses by scanning electron microscopy with an energy-dispersive X-ray analyzer (SEM-EDX) and atom probe tomography (APT) showed that a uniform Pt/Ir alloy was exposed on the tip surface as a clean surface without O or Cl contamination. The STM imaging using the fabricated tip showed that it is more suitable for investigating rough surfaces than conventional as-cut tips and applicable for atomic-resolution imaging. Furthermore, we applied the fabricated tip to qPlus AFM analysis in liquid and showed that it has atomic resolution in both the horizontal and vertical directions. Therefore, it is concluded that the amplitude-modulated AC etching method reproducibly provides sharp STM/AFM tips capable of both atomic resolution and large-area analyses without complex etching setups.
\end{abstract}

\maketitle

A platinum/iridium (Pt/Ir) alloy tip is widely used in scanning probe microscopy (SPM) because it has high mechanical strength and chemical stability, and an adequately sharp cut end can be easily obtained by mechanical cutting.\cite{Garnaes1990,Bian2021,Musselman1990} In particular, mechanically cut Pt/Ir tips are the first choice for high-resolution scanning tunneling microscopy (STM)\cite{Binnig1982} in ambient air, requiring a tip that does not react with air and is highly strong.\cite{Bian2021,Musselman1990} However, the tip shape of the mechanically cut wire is not reproducible\cite{Musselman1990} and often produces image artifacts, such as multiple-tip effects.\cite{Rogers2000,Gorbunov1993,Fotino1993} Therefore, a method to fabricate a sharp Pt/Ir tip with a controlled shape and high reproducibility is demanded.
Electropolishing is a low-cost and quick method for fabricating SPM tips.\cite{Melmed1991, Gewirth1989} Pt/Ir wires can be electropolished by the alternating-current (AC) electrolysis technique in \ce{CaCl2}\cite{Libioulle1995,Musselman1990,Zhang2020} or \ce{KCl}\cite{Takami2019} aqueous solution, typically at an AC voltage of $\SI{10}{V_{rms}}$ or higher\cite{Takami2019}. However, it is known to be difficult to obtain sufficiently sharp tips by this method; therefore, additional complex techniques, such as upside-down electrolytic baths (reversed-etching)\cite{Fotino1992,Fotino1993,Meza2015} or multi-step etching with different electrolytes\cite{Libioulle1995,Lindahl1998} and applied voltages\cite{Meza2015,Sorensen1999,Aoyama2017}, are often used together. The reversed-etching method is typically performed by inserting a wire from below into an electrolyte supported by a glass pipette,\cite{Meza2015} which requires a precise positioning technique combined with a high-quality vibration-isolation system.\cite{Meza2015} Also, in the multi-step techniques, the tip shape strongly depends on the position of the tip on the solution surface and when to stop the first-step etching and proceed to the following step, which depends on the operator's skills. Therefore, the reproducible electropolishing method for fabricating sharp Pt/Ir tips without a complex setup or procedure has been required.

AC electropolishing of Pt in the presence of \ce{Cl-} ions is explained by the oxidation of Pt and the mechanical removal of deposits on the tip surface by gas bubbles generated by the counter-reaction, as described by the following reaction formulas:\cite{Shrestha2014}
\begin{align}
 \mathrm{Cathode}&:\ce{2H+ + 2e- -> H2}\uparrow\\
 \mathrm{Anode}&:\ce{Pt + 4Cl- -> [PtCl4]^{2-} + 2e-},\label{eq:Anode1}\\
 &\ \ce{[PtCl4]^{2-} + 2Cl- -> [PtCl6]^{2-} + 2e-}\label{eq:Anode2}
\end{align}
During the reactions (\ref{eq:Anode1}) and (\ref{eq:Anode2}), \ce{PtO_x} is produced as an intermediate\cite{Shrestha2014}, which would lead to the formation of an oxide film on the tip surface. Also, \ce{[PtCl4]^{2-}} and \ce{[PtCl6]^{2-}} ions produced in the anodic reaction are easily precipitated as \ce{Pt} particles or platinum chloride \ce{PtCl_x} \cite{Takami2019} and often contaminate the tip surface. Therefore, mechanical exfoliation and flushing of these deposits from the tip surface by the electrochemically-generated gas bubbles plays a crucial role in the etching procedure as well as the anodization of Pt. However, the flushing rate by gas bubbles is usually insufficient against the anodization rate of Pt. Hence, a tip with an irregular shape and contamination on the surface is often produced in conventional AC etching.

To overcome this issue, we propose an approach that increases the generation rate of the gas bubbles against the anodization rate of Pt by modulating the amplitude of the AC voltage applied between the wire and the counter electrode. This approach is based on the fact that the oxidation of Pt does not proceed at voltages below about $\SI{14}{V_{0-p}}\ (\sim\SI{10}{V_{rms}})$, and alternatively, only the generation of \ce{H2} gas in equation (1) and the generation of \ce{Cl2} gas expressed in the following equation\cite{Patil2011} proceed.
\begin{align}
 \mathrm{Anode}&:\ce{2Cl- -> Cl2} \uparrow \ce{+ 2e-}
\end{align}
By modulating the amplitude of the applied voltage in AC electrolysis across $\SI{14}{V_{0-p}}$, the etching reaction with Pt oxidation and \ce{H2} gas generation above $\SI{14}{V_{0-p}}$ and the gas bubble generation of \ce{H2} and \ce{Cl2} without Pt anodization below $\SI{14}{V_{0-p}}$ are periodically repeated. Note that flushing with low-voltage-generated gases is commonly used as an independent step in the multi-step method to clean the tip surface after AC etching.\cite{Aoyama2017}
However, deposits produced during the etching affect the etching process and make the tip shape irregular. Therefore, removal of deposits by flushing after the etching process is not sufficient, and it is important to repeatedly remove the deposits by gas bubbles during the etching process.
Hence, the amplitude-modulation method is more straightforward than the conventional multi-step method and has the potential to produce sharper and cleaner tips than the multi-step method. In order to clarify whether this method can fabricate clean and sharp tips as opposed to conventional constant-amplitude AC electropolishing, we investigated the shape and chemical composition of the fabricated tips by scanning electron microscopy with energy-dispersive X-ray analyzer (SEM-EDX) and atom probe tomography (APT)\cite{Devaraj2018}. Furthermore, to evaluate the suitability for the various SPM analyses, we applied the fabricated tip to STM in air and atomic force microscopy (AFM)\cite{Binnig1986} in liquid and compared it with a conventional mechanically cut tip.

\begin{figure}[bt]
 \includegraphics[width=0.9\hsize]{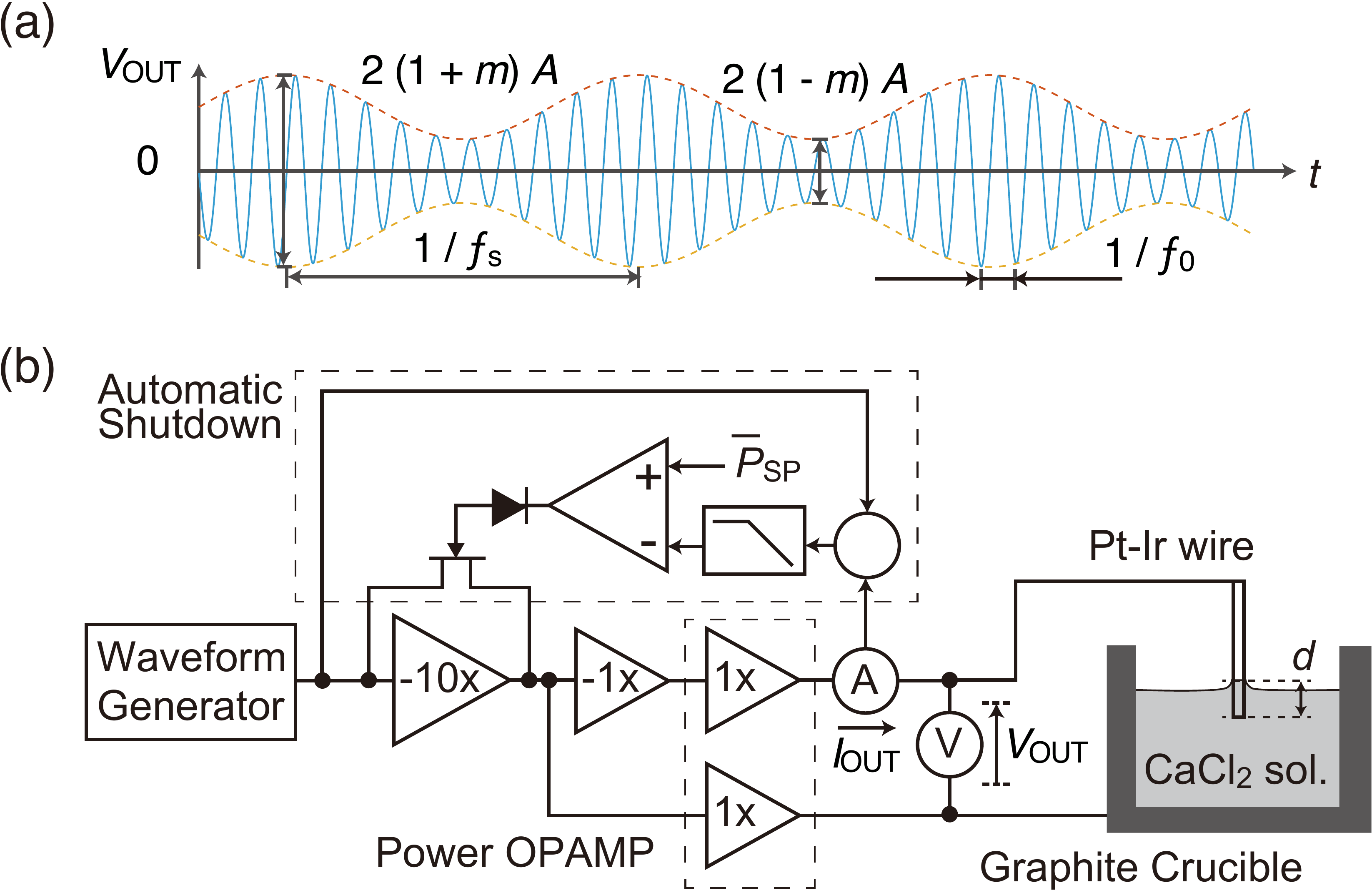}
 \caption{\label{fig:instruments} (a) An ideal waveform of the voltage applied to the cell $V_\mathrm{OUT}$ versus time $t$. (b) The block diagram of the setup for amplitude-modulated AC etching. }
\end{figure}

The pattern of applied voltage is shown in Figure \ref{fig:instruments}(a). This wave is obtained by modulating a fundamental sine wave of amplitude $A$ and frequency $f_0$ with a sine wave of frequency $f_s$ at modulation factor $m$. The following equation expresses the output voltage $V_\mathrm{out}$ versus time $t$.
\begin{align}
 V_\mathrm{out}(t)&=A\sin (2\pi f_0 t)\cdot \{1+m\sin (2\pi f_s t)\}
\end{align}
The values of $A$ and $m$ were determined as follows. First, the peak voltage $A(1+m)=\SI{27}{V}$ was chosen as close to the value $\SI{28}{V}\ (=\SI{20}{V_{rms}})$ of the previous study without amplitude modulation\cite{Libioulle1995,Aoyama2017,Takami2019}. Second, to maximize the amount of gas bubbles generation and to suppress the Pt oxidation, $A(1-m)=\SI{9}{V}$ was selected. To satisfy these relationships, $A=\SI{18}{V}$ and $m={50}{\%}$ were selected.

The amplitude-modulated voltage was generated by a waveform generator (AFG3022B; Tectronix Inc.) and amplified by homemade equipment. Figure \ref{fig:instruments}(b) shows a schematic of the setup. This circuit was driven by a $\pm\SI{15}{V}$ voltages supply and ideally provides differential voltage up to $\pm\SI{30}{V}$. Note that high output current $I_\mathrm{OUT}$ limits the maximum output voltages. Therefore, the waveform of both $V_\mathrm{OUT}$ and $I_\mathrm{OUT}$ were monitored by an independent circuit. The circuit also has the automatic shutdown section described in Supporting Information S1.

The experimental procedure was as follows. The acetone-saturated \ce{CaCl2} solution proposed by Libioulle \textit{et al.}\cite{Libioulle1995} was used as the electrolyte solution. \SI{40}{mL} of ultrapure water (UPW, $>\SI{18.2}{M\ohm\ m^{-1}}$) and \SI{40}{mL} of acetone ($>\SI{99.5}{\%}$, Nacalai Tesque Inc.) were mixed thoroughly, and \SI{14}{g} of \ce{CaCl2} anhydride ($>\SI{95.0}{\%}$, Nacalai Tesque Inc.) was added to the mixture. When \ce{CaCl2} was added, the mixture spontaneously separated into two layers, and the water-rich lower layer (namely acetone-saturated \ce{CaCl2} aqueous solution) was pipetted out and used as the electrolyte solution. \SI{10}{mL} of the electrolyte solution was poured into a \SI{25}{mL} graphite crucible, which was used as a counter electrode and a reaction cell. The platinum-iridium wire (Pt $\SI{80}{wt\%}-$Ir \SI{20}{wt\%}, Nilaco Corp.) with a diameter of \SI{0.1}{mm} was immersed in the electrolyte at a depth of $\sim\SI{1}{mm}$ utilizing the wire holder of commercial tungsten STM tip etcher (UTE-1001; Unisoku Inc.), which provides coarse positioning of the tip and the electrical connection to the tip. After the etching procedure, the fabricated tips were rinsed with UPW and ethanol ($\geq\SI{99.5}{\%}$, Nacalai Tesque Inc.) for \SI{5}{s} each at room temperature.

The fabricated tips were investigated using a field emission (FE-) SEM (JSM-6500F; JEOL Inc.) equipped with EDX (JED-2300F; JEOL Inc.). Elemental distribution on the tip surface was also analyzed by APT using a local-electrode-type\cite{Kelly2000} instrument (Ametek, LEAP4000X HR) in voltage mode at the temperature of 50 K.

STM analysis was performed on cleaved HOPG (Highly oriented pyrolytic graphite, grade ZYB; purchased from Alliance Biosystems, Inc.) and Pt-deposited mica substrates in the air at room temperature. Pt (\SI{99.99}{\%}; purchased from Furuuchi Chemical Corp.) was deposited on the cleaved surface of muscovite mica \ce{[KAl2(Si3Al)O10(OH)2]} substrates (purchased from Furuuchi Chemical Corp.) by electron beam evaporation in vacuum below \SI{1e-4}{Pa} at room temperature. 
AFM analysis was performed using a qPlus sensor\cite{Giessibl2019,Auer2023,Ichii2012,Ichii2014} built with a quartz tuning fork (QTF; purchased from SII Crystal Technology Inc.) and a fabricated Pt/Ir tip. The etched tip was cut into a length of $\sim\SI{1}{mm}$ and glued to the end of one prong of the QTF by epoxy resin (EPO-TEK H70E; Epoxy Technology Inc.), and the other prong was fixed to the mount. The KCl\{100\} surface in 1-methyl-1-propylpyrrolidinium bis(trifluoromethylsulfonyl)imide (Py$_{1,3}$-TF$_2$N) was used as the sample for the qPlus AFM analysis. The cleavage of the single crystal exposed the KCl\{100\} surface, and \SI{0.5}{\micro L} of Py$_{1,3}$-TF$_2$N (\SI{99.5}{\%}, purchased from IoLiTec GmbH) saturated with KCl ($>\SI{99.5}{\%}$, Fujifilm Wako Pure Chemical Corp.) was dropped on the surface and settled for \SI{24}{h} in the dry chamber before analysis. The STM and AFM analysis was performed by a system based on a commercial AFM/STM (JEOL Ltd., JSPM-5200) with a homebuilt STM head for small current detection and a homebuilt AFM head for qPlus sensor\cite{Ichii2012}. AFM analysis was performed in frequency modulation (FM) mode.\cite{Albrecht1991} The instrument's configuration was the same as previously reported.\cite{Bao2024,Nishiwaki2024}

\begin{figure}[bt]
 \includegraphics[width=0.7\hsize]{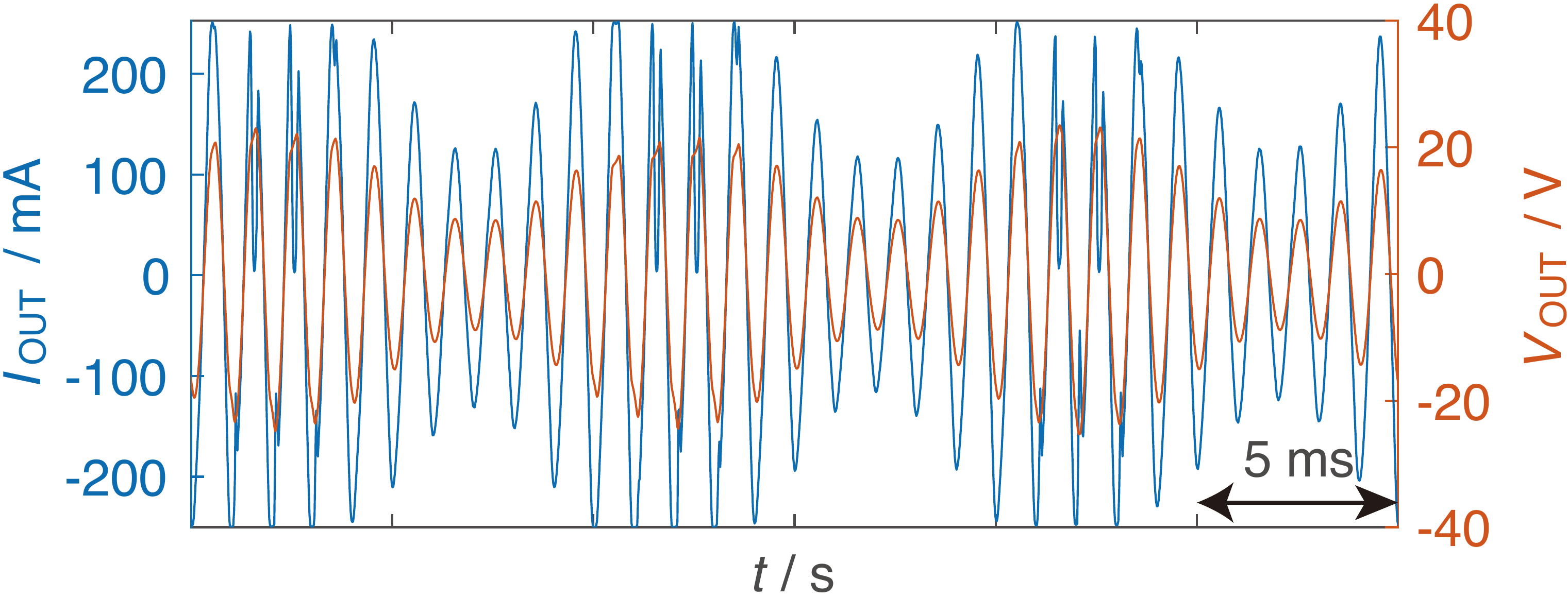} 
 \caption{\label{fig:IVcurves} The current $I_\mathrm{OUT}$ versus time $t$ curve obtained in the middle of the etching procedure. $f_0=\SI{1000}{Hz},\ f_s=\SI{100}{Hz},\ A=\SI{18}{V}$ and $m=\SI{50}{\%}$.}
\end{figure}

Figure \ref{fig:IVcurves} shows the current $I_\mathrm{OUT}$ and voltage $V_\mathrm{OUT}$ curves versus time $t$ obtained during the etching procedure at $f_0=\SI{1000}{Hz},\ f_s=\SI{100}{Hz},\ A=\SI{18}{V}$ and $m=\SI{50}{\%}$. $I_\mathrm{OUT}-t$ and $V_\mathrm{OUT}-t$ curves for the entire etching time and around the shutdown operation are shown in Figure S1 in the supporting information. 
In the region where $|V_\mathrm{OUT}|$ is approximately \SI{14}{V} ($\sim\SI{10}{V_{rms}}$) or higher, there is significant distortion in the $I_\mathrm{OUT}-t$ curve, while it is almost negligible in the region where $|V_\mathrm{OUT}|<\SI{14}{V}$. 
This suggests that the current was impeded by the bubbles or passivation film formation on the tip surface at high voltages of $|V_\mathrm{OUT}|>\SI{14}{V}$, which did not occur at $|V_\mathrm{OUT}|<\SI{14}{V}$. 
Note that, the $V_\mathrm{OUT}-t$ curve showed slight distortion in the region where the $I_\mathrm{OUT}$ showed the intense distortion. As mentioned above, a large $I_\mathrm{OUT}$ can limit $V_\mathrm{OUT}$, which may have distorted $V_\mathrm{OUT}-t$ curve. However, the distortion in $I_\mathrm{OUT}$ was obviously more significant than that in $V_\mathrm{OUT}$, and it is unlikely that the distortion in $I_\mathrm{OUT}$ was caused solely by the distortion in $V_\mathrm{OUT}$.
Considering that intense bubble generation was observed even when an amplitude $|V_\mathrm{OUT}|$ of \SI{14}{V} or less was applied, it is expected that anodic oxidation of Pt progressed only above \SI{14}{V}, and that only flushing by gas bubbles occurred below \SI{14}{V}.
Note that the passive film can be formed not only by the oxidation of Pt but also by the precipitation of Pt particles coated with impurities\cite{Takami2019} due to the reduction of the \ce{PtCl_x} and \ce{PtO_x}\cite{Shrestha2014} on the tip surface or the reduction of \ce{[PtCl4]^{2-}} and \ce{[PtCl6]^{2-}} ions dissolved in the electrolyte. Therefore, it is reasonable that the $I_\mathrm{OUT}-t$ curve shows distortions at high voltage in both anodic and cathodic reactions.

\begin{figure}[bt]
 \includegraphics[width=\hsize]{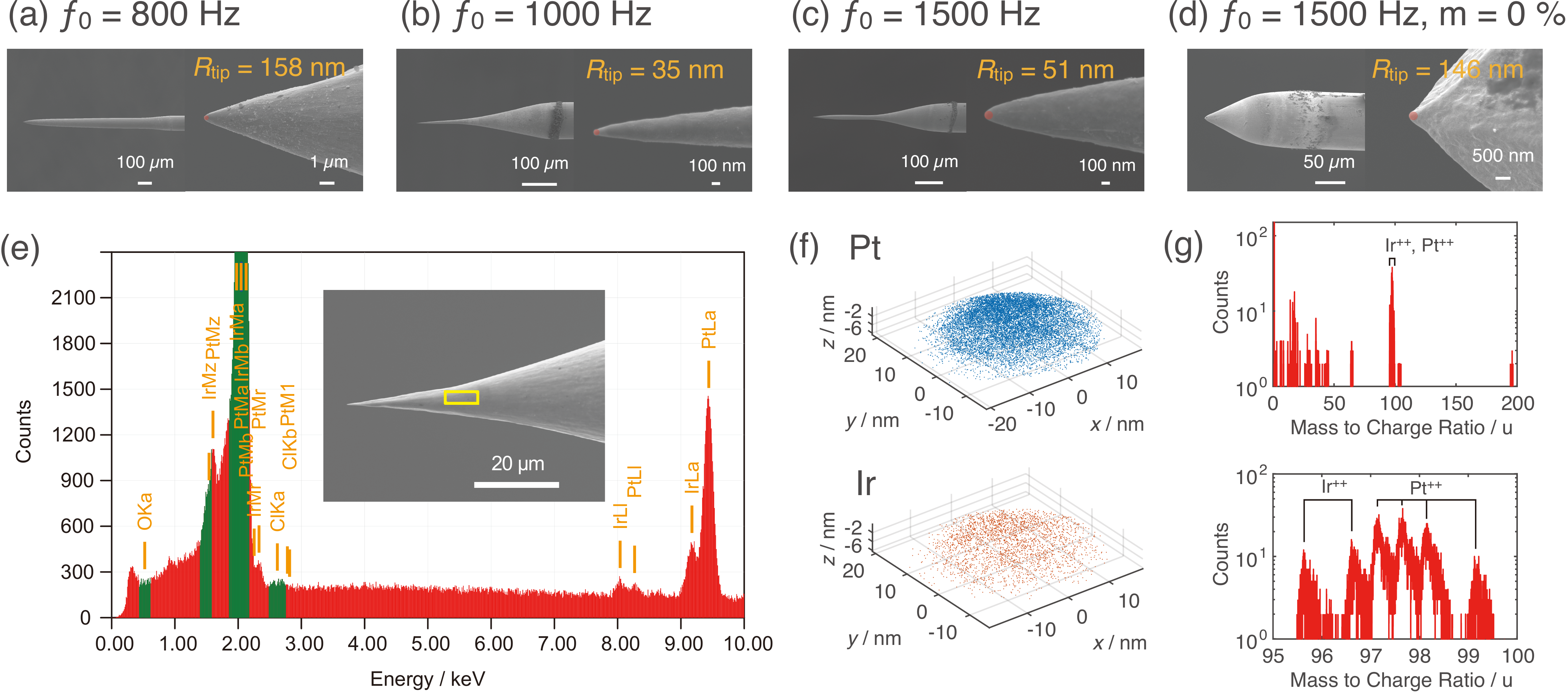} 
 \caption{\label{fig:Probe} (a-c) The SEM images of tips fabricated at $A=\SI{18}{V}$, $f_s=0.1f_0$, $m=\SI{50}{\%}$ and (a) $f_0=\SI{800}{Hz}$, (b) $f_0=\SI{1000}{Hz}$ and (c) $f_0=\SI{1500}{Hz}$, respectively. (d) The SEM image of a tip fabricated without amplitude modulation ($m=\SI{0}{\%}$) at $A=\SI{18}{V}$ and $f_0=\SI{1500}{Hz}$. (e) The EDX spectrum obtained at a tip surface fabricated at $A=\SI{18}{V},\ f_s=\SI{100}{Hz},\ m=\SI{50}{\%}$ and $f_0=\SI{1000}{Hz}$ in the region marked by a yellow box in the SEM image. (f, g) The distribution map of Pt and Ir atoms (f) and corresponding histogram of mass-to-charge ratio (g) obtained by APT analysis of a tip fabricated at $A=\SI{18}{V},\ f_s=\SI{100}{Hz},\ m=\SI{50}{\%}$ and $f_0=\SI{1000}{Hz}$.}
\end{figure}

Figure \ref{fig:Probe}(a-c) show SEM images of tips fabricated at $A=\SI{18}{V}$,  $f_s=0.1f_0$, $m=\SI{50}{\%}$ and (a) $f_0=\SI{800}{Hz}$, (b) $f_0=\SI{1000}{Hz}$ and (c) $f_0=\SI{1500}{Hz}$, respectively. Note that the $f_s$ value of $f_s=0.1f_0$ were selected based on the discussion in Supporting Information S2. At $f_0=\SI{800}{Hz}$ (Figure 3(a)), a tip with a long shank was fabricated, and its apex showed a relatively large opening angle and a radius of curvature of more than \SI{100}{nm}. In contrast, tips with a small opening angle and a radius of curvature of less than \SI{100}{nm} were fabricated at $f_0=\SI{1000}{Hz}$ and $f_0=\SI{1500}{Hz}$ (Figure 3(b, c)). As shown in Figure S3 of Supporting Information S2, the tips with a diameter of less than \SI{100}{nm} were fabricated at $\SI{900}{Hz}\leq f_0 \leq \SI{1500}{Hz}$ and $A=\SI{18}{V}$ without strong dependence on $f_0$, while the tip curvature radius and opening angle increased for $f_0\leq \SI{800}{Hz}$. The yields rate of the tips with a radius of \SI{100}{nm} at $f_0=\SI{1000}{Hz}$ is calculated to be \SI{67}{\%} as shown in Supporting Information S3.

We also fabricated a tip without amplitude modulation for comparison. Figure \ref{fig:Probe}(d) shows a tip fabricated without amplitude modulation at $A=\SI{18}{V}\ (=\SI{12.7}{V_\mathrm{rms}})$, $f_0=$1500 Hz, and $m=\SI{0}{\%}$. Although the tip radius was relatively small, its opening angle was obviously larger than the tip fabricated with amplitude modulation. We have experimented in detail for various $A$ and $f_0$, as shown in Figure S2 in Supporting Information S2, exploring the conditions to fabricate a sharp tip without amplitude modulation. Although relatively sharp tips with a radius of curvature of less than \SI{1}{\micro m} were obtained at $A=\SI{18}{V}$, most of them showed a distorted pyramid-like shape with a large opening angle, similar to Figure \ref{fig:Probe}(d). 
Also, the reproducibility of the tip shape and radius of curvature was limited, and many of them were covered with impurities or distorted in shape. In addition, tips with a radius of less than \SI{100}{nm} could only rarely be fabricated. The tip radius determines the spatial resolution, and the opening angle of the tip is crucial for investigating highly rough surfaces. In addition, tips with a small opening angle are beneficial in approaching the tip-to-tip distance in multi-probe SPM\cite{Voigtlander2018,Hasegawa2007} and suppressing long-range background interaction force in high-resolution AFM. Therefore, the amplitude modulation technique, which enables the fabrication of tips with small radius and opening angle, has advantages over the constant amplitude techniques.

Figure \ref{fig:Probe}(e) shows an EDX spectrum obtained at a tip surface. Hereafter, all analyses were performed with the tips fabricated by amplitude-modulated AC electropolishing with $f_0=\SI{1000}{Hz},\ f_s=\SI{100}{Hz},\ A=\SI{18}{V}$ and $m=\SI{50}{\%}$. Only Pt and Ir peaks were intense, and O and Cl peaks were negligibly small. Note that a C peak found at \SI{0.277}{keV} originates from carbon contamination \cite{Hugenschmidt2023} during the SEM analysis. This result indicates that no oxides or chlorides were detected on the tip surface, suggesting a clean metal surface was exposed. In addition, the distribution of Pt and Ir atoms on the tip surface was analyzed by APT. Figure \ref{fig:Probe}(f) shows the distribution maps for Pt and Ir atoms on the tip surface reconstructed from the detected positions and time of flight of the ionized atoms on the APT analysis.\cite{Devaraj2018} The result shows that the Pt and Ir elements are homogeneously distributed on the tip surface. Figure \ref{fig:Probe}(g) shows a histogram of mass-to-charge ratios corresponding to the areas analyzed in Figure \ref{fig:Probe}(f). The approximate elemental compositions estimated from the area of \ce{Pt^{++}} and \ce{Ir^{++}} peaks shown in Figure \ref{fig:Probe}(g) were \SI{83}{\%} for Pt and \SI{17}{\%} for Ir, which were close to those of the bulk.

\begin{figure}[bt]
 \includegraphics[width=\hsize]{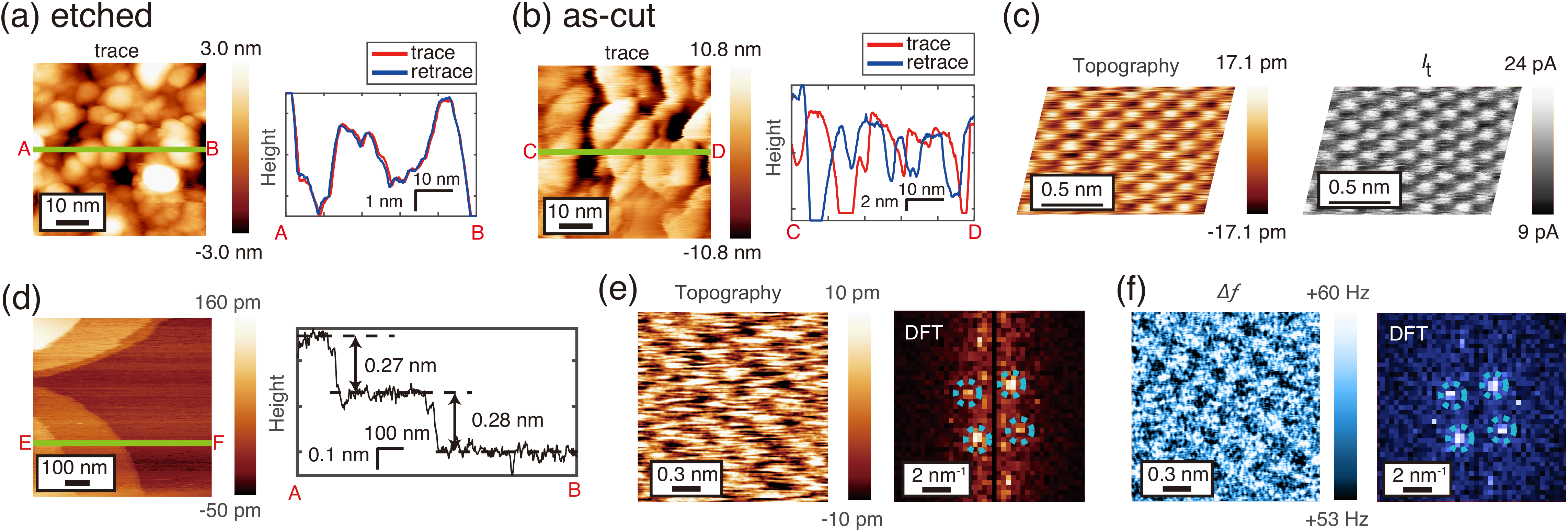} 
 \caption{\label{fig:SPM} (a, b) STM images and line profiles of Pt-deposited mica surface obtained using an (a) etched tip and (b) conventional as-cut wire. Bias Voltage$=511$ mV and tunneling current $I_\mathrm{t}\sim\SI{1e1}{pA}$. (c) STM topography ($z$) and tunneling current ($I_\mathrm{t}$) images of the HOPG surface obtained using an etched tip. Bias Voltage$=68$ mV. (d-f) qPlus AFM images obtained on KCl$\{100\}$ surface in KCl-saturated Py$_{1,3}$-TF$_2$N. (d) large-area topography with line profiles, (e) high-resolution topography with its discrete Fourier transform (DFT) image, and (f) corresponding $\Delta f$ image with its DFT image. $Q=145,\ f=\SI{18473}{Hz},\ A_\mathrm{p-p}\sim\SI{3e2}{pm}.$}
\end{figure}

Figure \ref{fig:SPM}(a) shows the STM topographic image of a Pt-deposited mica substrate obtained using an amplitude-modulated-AC-etched Pt/Ir tip. The spherical grains were clearly imaged, and the line profiles obtained on the trace and retrace scans were overlapped entirely, suggesting that a stable scan was performed. We also carried out STM imaging on the Pt substrate using an as-cut Pt/Ir wire without etching for comparison, as shown in Figure \ref{fig:SPM}(b). The deposited grains were imaged in a more distorted shape than those imaged by the etched tip, and the trace and retrace profiles showed a large discrepancy. This suggests that the irregular shape of the as-cut tip resulted in unstable scans and distorted images with the tip artifact. 
Therefore, 
although an as-cut tip has been applied to various STM analyses in air, including atomic-scale high-resolution analysis\cite{Pong2007}, an etched tip with a well-determined shape is especially advantageous for analyzing macroscopic features of highly rough surfaces. Note that simple multi-tip effects\cite{Chen2004} can also cause such grain-like patterns, and the trace/retrace profile discrepancy can also be caused by inappropriate distance feedback parameters during STM measurement as well as by tip deformation. A wide area scanned image, and its line profile are shown in the Supporting Information S4 to foreclose this possibility.

Also, a HOPG surface was imaged to demonstrate its applicability to high-resolution analysis. 
Figure \ref{fig:SPM}(c) shows the topography and tunneling current ($I_\mathrm{t}$) images obtained on the cleaved surface of HOPG. Both images show the well-known six-fold symmetry contrast which is consistent with the arrangement of $\beta$ carbon atoms.\cite{Wang2006,Schneir1986,Tomanek1987} This means that the etched tip provides a stable analysis even on the atomic scale and can replace the conventional as-cut tips in STM. 

Furthermore, the fabricated tips were also applied to high-resolution FM-AFM analysis. Figure \ref{fig:SPM}(d) shows a large-area topographic image of the KCl\{100\} surface and its line profile obtained by qPlus AFM in KCl-saturated Py$_{1,3}$-TF$_2$N. Atomically flat terraces and atomic steps with a height of $\sim\SI{0.27}{nm}$ were clearly imaged, where the obtained step height was close to the thickness of a single atomic layer on KCl\{100\} of \SI{0.31}{nm}\cite{Walker2004,Schwabegger2013}. Figures \ref{fig:SPM}(d) and \ref{fig:SPM}(e) show a high-resolution topographic image and the corresponding frequency-shift ($\Delta f$) image obtained on the terrace surface, respectively. Although the signal-to-noise ratio is poorer than that reported using an electropolished W tip, both images show four-fold symmetric contrasts corresponding to the arrangement of K or Cl ions in KCl\{100\} surface\cite{Schwabegger2013}, and the corresponding four bright spots are clearly shown in the discrete Fourier transform (DFT) image. This means that the resolution reaches the atomic scale for both the vertical and horizontal directions. 

In summary, we developed a method to fabricate sharp and clean Pt/Ir tips by amplitude-modulated AC electropolishing. The sharp tips with a radius of curvature less than $\SI{100}{nm}$ were obtained at $\SI{900}{Hz}\leq f_0\leq\SI{1500}{Hz}, \ f_s=0.1f_0,\ A=\SI{18}{V}$ and $m=\SI{50}{\%}$. SEM-EDX and APT analyses showed that a uniform Pt/Ir alloy was exposed on the tip surface as a clean surface without O or Cl contamination. The STM imaging using the fabricated tip showed that it is more suitable for investigating rough surfaces than as-cut tips and applicable for atomic-resolution imaging. Furthermore, we applied the fabricated tip to qPlus AFM analysis in liquid and showed that it has atomic resolution in both the horizontal and vertical directions. Therefore, it is concluded that the amplitude-modulated AC etching method reproducibly provides sharp STM/AFM tips capable of both atomic resolution and large-area analyses without complex etching setups.

\section*{Supplementary Material}

See the Supporting Information file for more details on the following data and supplementary discussions.
\begin{enumerate}[S1.]
  \item Control of applied voltage by automatic shutdown circuit.
  \item Etching parameter dependence of the tip shape and comparison with conventional method without amplitude modulation.
  \item Evaluation of the reproducibility of etched tips by consecutive fabrication.
  \item STM images of platinum-deposited mica substrates acquired over a wide area.
\end{enumerate}

\begin{acknowledgments}
  This work was supported by a Grant-in-Aid for Scientific Research B (JP23H01850) from the Japan Society for the Promotion of Science (JSPS).
\end{acknowledgments}

\section*{Author Declarations}
\subsection*{Conflict of Interest}
The authors have no conflicts to disclose.

\subsection*{Author Contributions}
Y.N. drafted the original paper. Y.N. developed the etching equipment and fabricated the sample tip. Y.N performed the STM, AFM, and SEM-EDX experiments. S.K. performed the APT experiment. T.I. developed the STM/AFM equipment and coordinated the project. T.U., S.K., and T.I. contributed to the interpretation of the results. All authors discussed the results and contributed to the preparation of the paper.

\section*{Data Availability Statement}

The data that support the findings of this study are available from the corresponding author upon reasonable request.

\bibliography{references}%

\end{document}


\preprint{}

\title{Supporting Information for ``One-step Fabrication of Sharp Platinum/Iridium Tips via Amplitude-Modulated Alternating-Current Electropolishing''}

\author{Yuto Nishiwaki}
\author{Toru Utsunomiya}
\author{Shu Kurokawa}
\author{Takashi Ichii*}%
 \email{ichii.takashi.2m@kyoto-u.ac.jp}
\affiliation{ 
  Department of Materials Science and Engineering, Kyoto University, Yoshida Honmachi, Sakyo, Kyoto, 606-8501, Japan. %
}%

\date{\today}%

\maketitle

\section{Control of applied voltage by automatic shutdown circuit}
\begin{figure}[bt]
  \includegraphics[width=0.7\hsize]{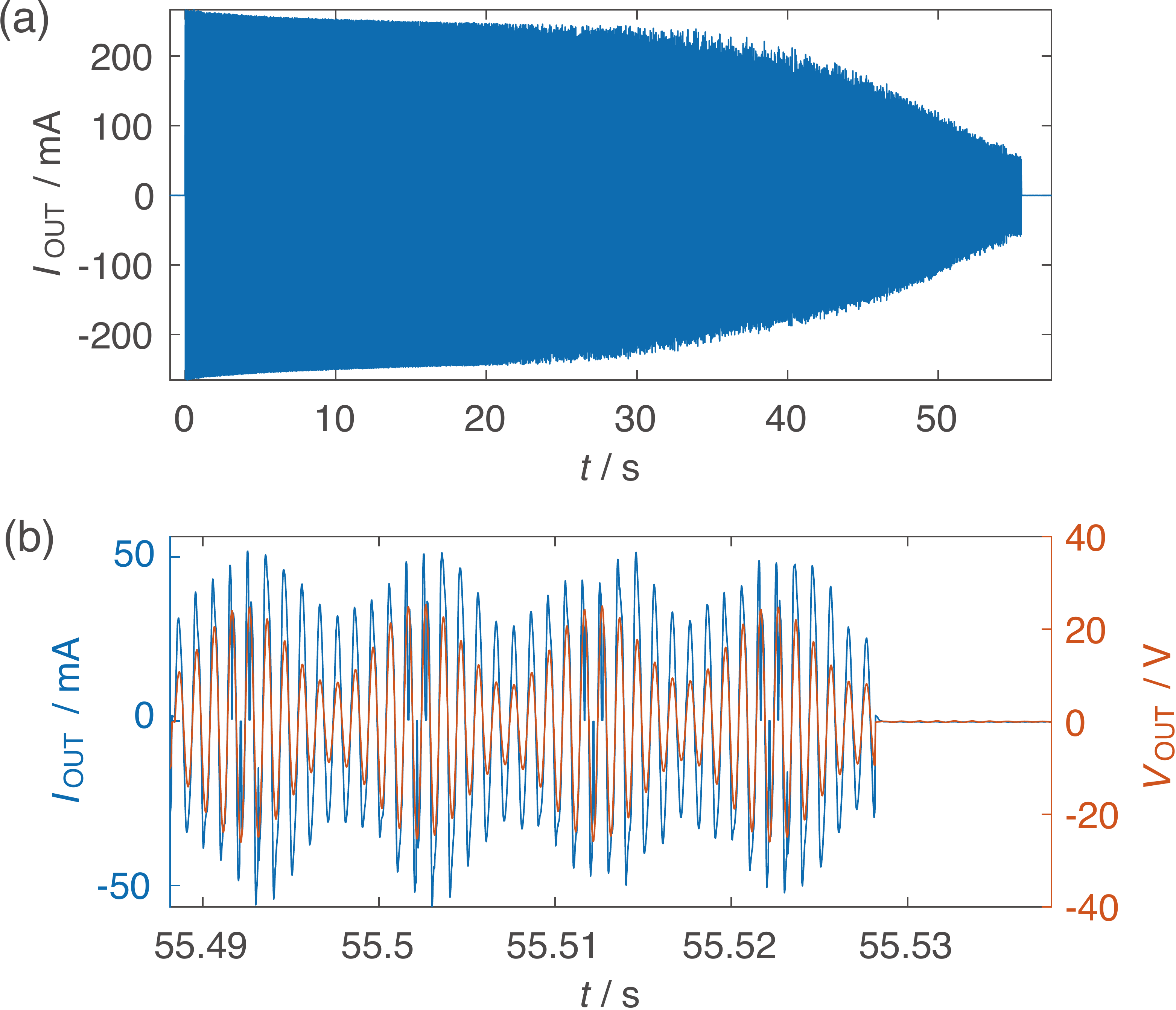}%
  \caption{\label{fig:SI_IVcurves} (a) The current $I_\mathrm{OUT}$ versus time $t$ curve obtained for entire etching time. (b) The current $I_\mathrm{OUT}$ and voltage $V_\mathrm{OUT}$ versus time $t$ curves obtained at the end of the etching procedure.$\ f_0=\SI{1000}{Hz},\ f_s=\SI{100}{Hz},\  A=\SI{18}{V}$ and $m=\SI{50}{\%}$.}
\end{figure}

Figure \ref{fig:SI_IVcurves} shows the current $I_\mathrm{OUT}$ and voltage $V_\mathrm{OUT}$ curves versus time $t$ obtained during etching procedure at $f_0=\SI{1000}{Hz},\ f_s=\SI{100}{Hz},\ A=\SI{18}{V}$ and $m=\SI{50}{\%}$. Figure \ref{fig:SI_IVcurves}(a) shows the $I_\mathrm{OUT}-t$ curve for the entire etching time. $I_\mathrm{OUT}$ decreases during the etching procedure because the resistance increases as the etching and necking of the wire near the liquid surface progresses.\cite{Libioulle1995,Takami2019} Note that, even without a shutdown circuit, $I_\mathrm{OUT}$ and $\bar{P}$ drop off at the end as the tip is formed. However, an automatic shutdown circuit is required to suppress the voltage application after the drop-off of the current to prevent over-etching after tip formation and reduce the radius of the curvature of the tip, as mentioned in the main text.

The automatic shutdown circuit is widely used to prevent over-etching after the tip formation,\cite{Borzenets2012,Nakamura1999,Takami2019,Meza2015} which causes an increase in the tip radius.\cite{Nakamura1999} In this experiment, we employed an architecture using a field effect transistor (FET) to eliminate the error of tip shape caused by the mechanical relay's delay. The automatic shutdown circuit in Figure 1(b) detects the drop-off of the wire and completion of the etching procedure\cite{Meza2015,Takami2019} by monitoring the effective power consumed by electrolysis $\bar{P}$ and automatically shut down the voltage output at $\bar{P}_\mathrm{SP}$. 

The circuit shown in Figure 1(b) calculated the effective power $\bar{P}$ by smoothing the instantaneous power $P=I_\mathrm{OUT}\cdot V_\mathrm{OUT}$ with an 8th-order ($\SI{-48}{dB/oct}$) Bessel low-pass filter (MAX292; Analog Devices Inc.) with a cutoff frequency  $f_\mathrm{LPF}$. 
It shut down the voltage output when $\bar{P}$ became less than $\bar{P}_\mathrm{SP}\sim\SI{0.10}{W}$. Figure \ref{fig:SI_IVcurves}(b) shows the $I_\mathrm{OUT}$ and $V_\mathrm{OUT}$ versus $t$ curves at the end of the etching. The $V_\mathrm{OUT}$ was shut down at the end of etching in the fall time within 50 µs, which is sufficiently smaller than the operating time or the release time of a conventional mechanical relay. The voltage fluctuation after the shutdown was less than $\SI{0.3}{V_{p-p}}$, which was small enough to stop the anodization and gas generation reaction.

To effectively suppress over-etching, the internal delay time of the automatic shutdown circuit must be minimized. However, the bandwidth of the shutdown circuit is regulated by $f_\mathrm{LPF}$, which must be smaller than the frequency of harmonic components in $P$. For example, to obtain at least $\SI{-24}{dB/oct}$ attenuation for the first harmonic component ($=f_0$), it is necessary to set $f_\mathrm{LPF}<1/\sqrt{2} f_0\ (\sim 0.7 f_0)$. For this reason, $f_\mathrm{LPF}=\SI{709}{Hz}$ for $\SI{900}{Hz} \leq f_0 \leq \SI{1500}{Hz}$ and $f_\mathrm{LPF}=\SI{333}{Hz}$ for $\SI{650}{Hz} \leq f_0 < \SI{900}{Hz}$ was taken in this study.

\section{Etching parameter dependence of the tip shape and comparison with conventional method without amplitude modulation}
The dependence of tip shape on $f_0,\ f_s,\ A,\ m$ was investigated to find the etching parameters suitable for obtaining a sharp and clean tip. First, for comparison, etching at constant amplitude without modulation ($m=\SI{0}{\%}$, \textit{i.e.}, $V_\mathrm{OUT}=A \sin (2\pi f_0 t)$) was performed. Figure S2 shows SEM images of tips fabricated without amplitude modulation for each $f_0$ and $A$.
\begin{figure}[bt]
  \includegraphics[width=0.99\hsize]{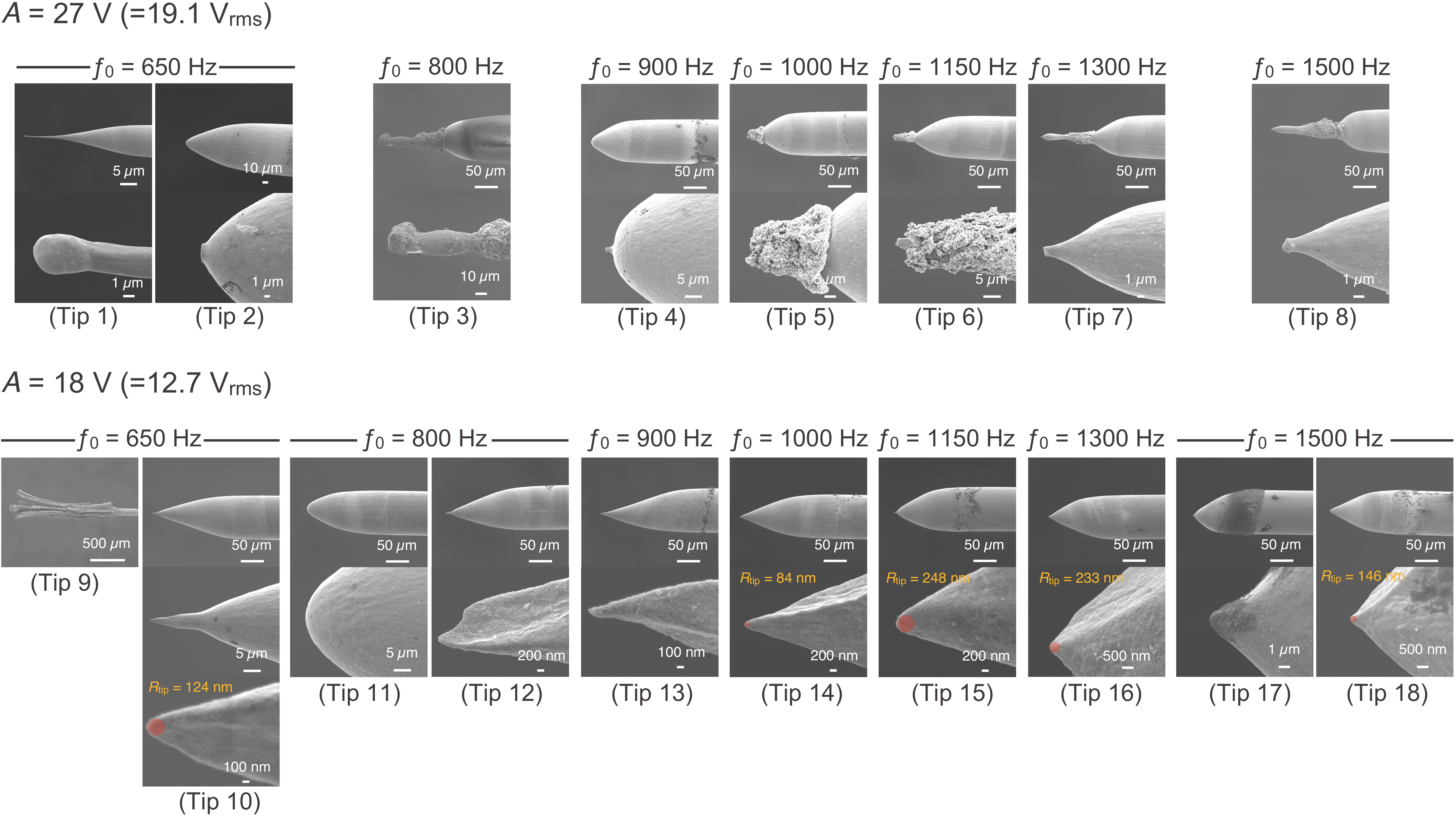}%
  \caption{\label{fig:SI_withoutAM_parameter} SEM images and radii of curvature $R_\mathrm{tip}$ of tips fabricated without amplitude modulation ($m=\SI{0}{\%}$) at each $f_0$ and $A$. $f_\mathrm{LPF}=\SI{709}{Hz}$ for $\SI{900}{Hz} \leq f_0 \leq \SI{1500}{Hz}$ and $f_\mathrm{LPF}=\SI{333}{Hz}$ for $\SI{650}{Hz} \leq f_0 < \SI{900}{Hz}$.}
\end{figure}
Tips 1-8 were fabricated with $A=\SI{27}{V}$, which was selected as the close value to the previous research\cite{Libioulle1995,Aoyama2017} of $A=\SI{28.2}{V}\ (=\SI{20}{V_{rms}})$. Although Tip 1 fabricated with $f_0=\SI{650}{Hz}$ appeared to be sharpened in the wide-area SEM image, a spherical object with a few micrometers in diameter was observed at the tip in the magnified image. Also, except for Tip 1, 2, and 4, porous deposits were formed at the neck or the end of the tip. A clean tip apex was exposed in Tip 7 and Tip 8, though they had the porous deposits on its neck and a distorted tip shape. These results imply that the obstruction of the etching process due to the deposition of impurities on the tip could have resulted in a distorted tip shape. Even in Tip 1, 2, and 4 without significant contamination, it is possible that impurities caused the distorted tip shape, and then the impurities detached from the tip surface due to gas bubble flushing. %

The distorted tip shape is also reported at $A=\SI{28.2}{V}$ and $f_0=\SI{60}{Hz}$,\cite{Libioulle1995} \SI{100}{Hz},\cite{Aoyama2017} and \SI{1000}{Hz}.\cite{Aoyama2017} Our results in $A=\SI{27}{V}$ are reasonable since it is known that applying AC voltage with an amplitude much larger than $A=\SI{14.1}{V}\ (=\SI{10}{V_{rms}})$ tends to roughen and distort the tip shape\cite{Sorensen1999}. However, such distorted tips were also fabricated with $A=\SI{18}{V}$, as clearly shown in Tip 12-14, 16, and 18, which showed a pyramid-like shape  with large opening angles. %
In addition, tips completely covered with impurities (Tip 9 and Tip 17) and tips with extremely large radius of curvature (Tip 11) were also produced. %
Therefore, the simple method without amplitude modulation is insufficient to
reproducibly fabricate clean tips with small opening angles and radii.

Next, etching with the amplitude-modulation technique was performed. From the principle of the amplitude modulation method presented in the main text, the minimum value of the applied voltage amplitude (\textit{i.e.}, $A(1-m)$) should be large enough to maximize the bubble generation rate and prevent progressive oxidation of Pt. Without amplitude modulation ($m=0$), we confirmed that Pt was slightly oxidized at $A=\SI{13.5}{V}$ and that the bubble was generated without the Pt oxidation at $A=\SI{9.0}{V}$. Thus, we used a pair of $A$ and $m$ such that $A(1-m)=\SI{9.0}{V}$ in the experiment. Similarly, $f_s$ must be sufficiently large to prevent irregular etching of the tip due to deposits produced during the etching. However, as shown in Figure 2 in the main text, a significant distortion of the $I_\mathrm{OUT}-t$ curve appears only for two periods of the carrier wave ($f_0$) during one period of the signal wave ($f_s$) when $f_0=\SI{1000}{Hz}$ and $f_s=0.1 f_0$. As discussed in the main text, the distortion of the $I_\mathrm{OUT}-t$ curve suggests the formation of a passive layer due to oxidation of Pt. Therefore, Pt was mainly oxidized when the $I_\mathrm{OUT}$ was heavily distorted. Since the $I_\mathrm{OUT}$ was found to be heavily distorted only twice for a single amplitude modulation, $f_s/f_0$ would be at most 0.1 to sufficiently oxidize Pt. Therefore, $f_s$ was fixed to $0.1f_0$ each in the experiment. 
\begin{figure}[bt]
  \includegraphics[width=0.99\hsize]{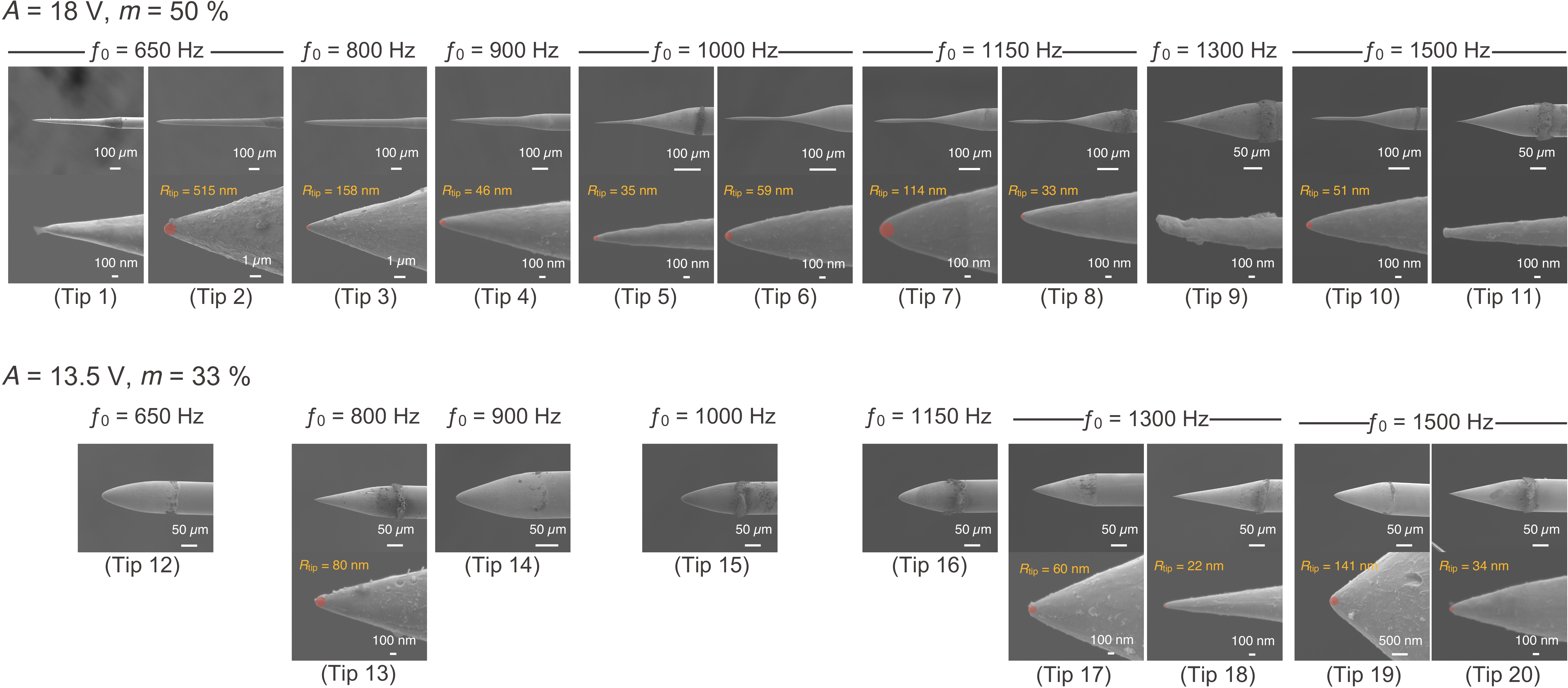}%
  \caption{\label{fig:SI_AM_parameter} SEM images and radii of curvature $R_\mathrm{tip}$ of tips fabricated by amplitude-modulated voltage at each $f_0$ and $A$. $f_s$ was fixed to $0.1f_0$. $m$ was chosen based on $A$ by the relationship of $A(1-m)=\SI{9}{V}$ described in the text. $f_\mathrm{LPF}=\SI{709}{Hz}$ for $\SI{900}{Hz} \leq f_0 \leq \SI{1500}{Hz}$ and $f_\mathrm{LPF}=\SI{333}{Hz}$ for $\SI{650}{Hz} \leq f_0 < \SI{900}{Hz}$.}
\end{figure}

Figure \ref{fig:SI_AM_parameter} shows SEM images of the tips fabricated by the amplitude-modulated voltage in each $A$ and $f_0$ with the derived $m$ and $f_s$. For $A=\SI{18}{V}$ ($m = \SI{50}{\%}$; \textit{i.e.,} $A(1-m)=\SI{9}{V}$ and $A(1+m)=\SI{27}{V}$), cone-shaped tips with a relatively small opening angle was obtained for all $f_0$. The tip apex was clean and smooth except for Tip 1, 9, and 11, and the radius of curvature was less than 100 nm at $f_0\geq\SI{900}{Hz}$ except for Tip 7, 9, and 11. That is, over a wide range of $\SI{900}{Hz}\leq f_0\leq \SI{1500}{Hz}$, tips with a radius less than 100 nm can be obtained with an average yield of \SI{63}{\%}. This yield is in good agreement with the yield for only $f_0=\SI{1000}{Hz}$ shown in Supporting Information S3 (see below), implying that there is no significant $f_0$ dependence of the tip radius of curvature within the range of $\SI{900}{Hz}\leq f_0\leq \SI{1500}{Hz}$ at $A=\SI{18}{V}$. In contrast, in the $f_0\leq\SI{800}{Hz}$ region, the tips showed a larger tip radius than 100 nm except Tip 1, which showed a distorted tip apex. As described in Supporting Information S1, when etching at low $f_0$, the cutoff frequency of LPF ($f_\mathrm{LPF}$) was also set low. This would be one cause of the increase in tip radius. Also, as $f_0$ is decreased, $f_s$ must also be decreased. Thus, it is not easy to reproducibly obtain a sharp tip with $A=\SI{18}{V}$ and $f_0<\SI{900}{Hz}$ when either low $f_0$ or $f_\mathrm{LPF}$ is increasing the tip radius. 

For $A=\SI{13.5}{V}$ ($m = \SI{33}{\%}$; \textit{i.e.,} $A(1-m)=\SI{9}{V}$ and $A(1+m)=\SI{18}{V}$), clean  tips without significant contamination were also produced for all tips. However, the tips fabricated at $f_0\leq \SI{1150}{Hz}$ showed large tip radius over \SI{1}{\micro m}, except for Tip 13. Such tips with a radius of \SI{1}{\micro m} or larger were also fabricated when the amplitude was not modulated (Tip 10 in Figure S2). For $f_0\geq\SI{1300}{Hz}$, relatively small tip diameters were obtained. However, some of the tips showed a significantly large opening angle, which was similar to the case without amplitude modulation.
From these results, we conclude that a tip with a small radius of curvature and opening angle can be fabricated under $A=\SI{18}{V}$ and $\SI{900}{Hz}\leq f_0\leq \SI{1500}{Hz}$ with the derived parameters $m=\SI{50}{\%}$ and $f_s=0.1 f_0$.

\section{Evaluation of the reproducibility of etched tips by consecutive fabrication}
\begin{figure}[bt]
    \includegraphics[width=0.99\hsize]{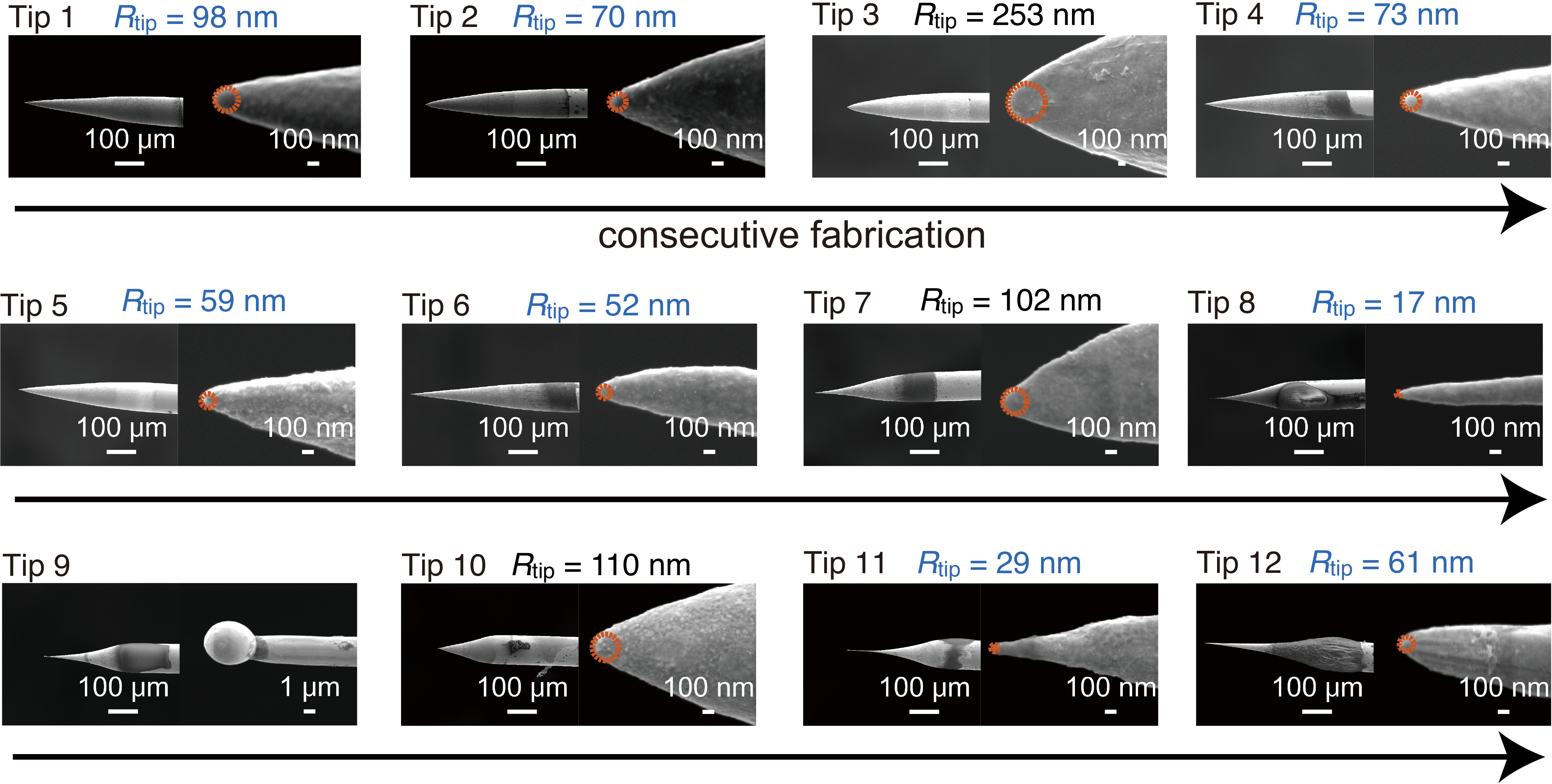}%
    \caption{\label{fig:SI_continuous} SEM images and radii of curvature $R_\mathrm{tip}$ of a series of tips fabricated in a single bath at $A=\SI{18}{V}$, $f_0= \SI{1000}{Hz}$, $m=\SI{50}{\%}$ and $f_s=0.1 f_0$.}
  \end{figure}

Figure \ref{fig:SI_continuous} shows SEM images and radius of curvature $R_\mathrm{tip}$ of a series of tips fabricated in a single bath at $A=\SI{18}{V}$, $f_0= \SI{1000}{Hz}$, $m=\SI{50}{\%}$ and $f_s=0.1 f_0$. Except for Tip 9, the tips have a radius of curvature less than 300 nm, of which 67 \% have a radius less than 100 nm. The round shape in Tip 9 may have been caused by the melting by Joule heating,\cite{Sorensen1999} which was sometimes observed regardless of the times the bath was used. On the surface of Tip 12, despite the small radius of curvature, a black precipitate was deposited on the shoulder of the tip shank, which appeared as a shadow on the SEM image. At this time, the bath was contaminated by black dispersed particles of Pt and \ce{PtCl_x}, and the viscosity was increased by the evaporation of acetone, which probably prevented sufficient removal of the deposits on the tip surface. Therefore, the lifetime of the electrolyte is limited to approximately 12 cycles, within which a clean tip with a radius of curvature of less than 100 nm can be obtained with a production yield rate of approximately 67 \%.
\clearpage

\section{STM images of Platinum-deposited mica substrates acquired over a wide area}
Figures 4(a) and (b) in the main text show STM images of a Pt-deposited mica substrate obtained with an etched and as-cut tip. The scan range of Figures 4(a) and (b) was 50 nm $\times$ 50 nm, which was comparable to the radii of tips. In such situations, the multi-tip effect can also cause similar STM images with multiple grains. Thus, to eliminate this concern, we obtained STM images of the larger area of the same places with Figures 4(a) and (b), as shown in Figure S5. The trace/retrace images were almost identical for the etched tip, and the line profiles overlapped well, as shown in Figure 4(a). The scan range of Figure S5 (300 nm $\times$ 300 nm) is larger enough than the tip radius ($<$ 100 nm). Therefore, these grain-like STM images are not considered to be caused solely by the multi-tip effect, although the contribution of the multi-tip effect cannot be completely foreclosed in the scan of such grain-like substrates.
\begin{figure}[bt]
  \includegraphics[width=0.99\hsize]{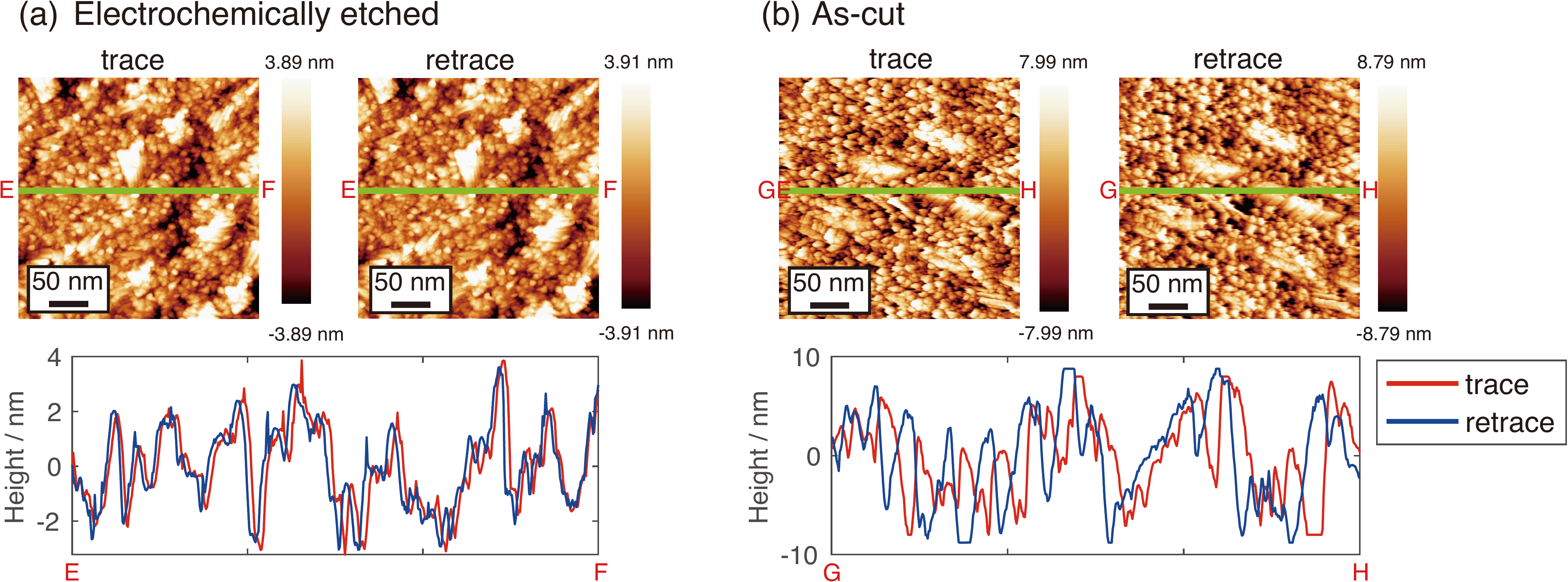}%
  \caption{\label{fig:SI_wideSTM} STM images and line profiles of Pt-deposited mica surface obtained in the range of $\SI{300}{nm}\times\SI{300}{nm}$ during the experiments in Figure 4(a, b). (a) Etched tip and (b) conventional as-cut tip were used. Bias Voltage$=511$ mV and tunneling current $I_\mathrm{t}\sim1\times10^1$ pA. The etched tip (a) was fabricated by amplitude-modulated voltage at $A=\SI{18}{V}$, $f_0= \SI{1000}{Hz}$, $m=\SI{50}{\%}$ and $f_s=0.1 f_0$.}
\end{figure}

In addition, in Figure 4(b) and Figure S5(b) obtained by the as-cut tip, a large discrepancy of the line profiles was observed between the trace and retrace scans. The discrepancy can be caused by inappropriate distance feedback parameters in the STM analyses as well as by the tip shape. However, some correspondence in the trace/retrace line profiles can be recognized in Figure S5(b), while such a correspondence cannot found in Figure 4(b). The correspondence accompanied by hysteresis in the trace/retrace profiles of the large-area scan indicates that it is more likely due to changes in the tip shape rather than the oscillation or feedback error due to inappropriate feedback parameters.

It should also be noted that the as-cut tip showed a larger height deviation than the etched tip in both Figures 4(a), (b) and Figure S5. Since we optimized the feedback parameters during the STM imaging for each image, they are not strictly identical. However, it is reasonable to assume that the difference in the tip shape caused such a height difference rather than the slight difference in feedback parameters. This is because the etched tip has a smooth spherical end, whereas the as-cut tip has many mini-tips on its rough end. If mini-tips are present on the tip end, the diameter of the mini-tip just in contact with the substrate determines the apparent radius of curvature of the tip. This may have resulted in a larger height difference in the STM image obtained with an as-cut tip.

\bibliography{references}%